\documentclass[12pt]{iopart}
\usepackage{graphicx}

\begin{document}
\title[Revealing virtual processes...]{Revealing virtual processes of a quantum Brownian particle in the phase space}

\author{S Maniscalco}

\address{\dag\ School of Pure and Applied Physics, University of KwaZulu-Natal, 4041 Durban, South Africa.}

\address{\ddag\ INFM, MIUR and Dipartimento di Scienze Fisiche ed
Astronomiche dell'Universit\`{a} di Palermo, via Archirafi 36, 90123
Palermo, Italy.}

\ead{maniscalco@ukzn.ac.za}

\begin{abstract}
The short time dynamics of a quantum Brownian particle in a harmonic
potential is studied in the phase space. An exact non-Markovian
analytic approach to calculate the time evolution of the Wigner
function is presented. The dynamics of the Wigner function of an
initially squeezed state is analyzed. It is shown that virtual
exchanges of energy between the particle and the reservoir,
characterizing the non-Lindblad short time dynamics where
system-reservoir correlations are not negligible, show up in the
phase space.
\end{abstract}

\submitto{\JOB} 

\maketitle

\section{Introduction}

During the last decades applications based on the most radical
predictions and features of quantum mechanics, namely the existence
of quantum superpositions and the entanglement, have been conceived
\cite{nielsen}.  Both of these features lie at the root of a new
class of quantum technologies exploiting the coherent manipulation
of quantum systems. Such process involves very delicate procedures
since the inevitable interaction of  the quantum systems with their
surrounding leads to loss of information and causes two types of
phenomena countering the coherent control of quantum systems:
decoherence and dissipation. Decoherence, which is the loss of phase
coherence between superpositions of quantum states, and dissipation,
which is the leakage of population from the system to the
environment, are the major enemies of  quantum technologies. In
order for quantum technologies to emerge, decoherence and
dissipation need to be defeated or controlled, and for that they
must be completely understood.

The theory of open quantum systems deals with the microscopic
description of quantum  systems interacting with their surroundings
\cite{petruccionebook}. It is of crucial importance to incorporate
methods and tools of the theory of open quantum systems to the
investigation of quantum technologies. Understanding and being able
to describe the effects of the environment on quantized systems is
the pre-requisite for the implementation of realistic applications.

The exact analytic description of the dynamics of an open system is
not an easy task. Commonly a series of approximations are necessary
in order to obtain treatable equations of motions for the density
matrix of the open system. Typical approximations are the weak
coupling and the Markovian approximations. The first one assumes
that the coupling between the system and the environment is weak
while the second one assumes that the correlations between system
and environment are negligible.

The miniaturization process which has lead to the possibility of
manipulating coherently quantum systems, exploiting the crucial
features of quantum theory, has been accompanied by the need of
speeding up the quantum engineering methods. This is because the
quantum properties of the systems, such as entanglement, gets
eventually destroyed by the interaction with the environment. The
description of the short time dynamics of an open quantum system
requires exact approaches which do not rely on the Markovian
approximation. Recently, a non-Markovian description of quantum
computing showing the limits of the Markovian approach has been
presented \cite{Alicki02}. Moreover, for many solid state systems,
the properties of the reservoirs are such that the Markovian
assumption does not hold at any time. This is e.g. the case of atom
lasers and photonic band gap materials \cite{fotolaser}. Therefore
the study of the dynamics of paradigmatic open quantum systems using
non-Markovian analytical approaches is crucial for the development
of prototypes of quantum devices which are the building blocks of
tomorrow's technologies.

In this paper I focus on an ubiquitous open quantum system, namely
the damped harmonic oscillator or quantum Brownian particle in a
harmonic potential \cite{QBMpapers}. Recently an analytic approach
for solving the non-Markovian master equation for this system via
the simmetrically ordered characteristic function has been
presented \cite{PRAsolanalitica}, and its dynamics has been
studied \cite{Maniscalco04b}. One of the new dynamical features
discovered in Ref. \cite{Maniscalco04b} is the existence of a
regime characterized by virtual exchanges of energy between the
system oscillator and its surrounding. Here we focus on this
regime and we present a method to calculate exactly the
non-Markovian time evolution of the Wigner function of an
initially squeezed state of the oscillator. It has been
demonstrated in Ref. \cite{Maniscalco04a} that the regime
considered in this paper may be experimentally simulated with
single trapped ions. It is worth recalling that, for these
systems, measurements of the Wigner function have been performed
in the experiments \cite{Monroe96}.

The paper is structured as follows. In Sec.\ref{sec:system} we
review the properties of the system, introducing the master
equation and its solution. Section \ref{sec:wigner} contains the
main result of the paper, namely the derivation of the exact
dynamics of the Wigner function for an initially squeezed state.
Finally Sec. \ref{sec:conclusions} presents conclusions.

\section{The system and the master equation}\label{sec:system}
The reduced density matrix $\rho_S$ describing the dynamics of the
quantum Brownian particle obeys, in the secular approximation, to
the following non-Markovian master equation \cite{EPJRWA}
\begin{eqnarray}
\frac{ d \rho_S}{d t}= &-&  \frac{\Delta(t) + \gamma (t)}{2} \left[
a^{\dag} a \rho_S - 2 a \rho_S a^{\dag} + \rho_S a^{\dag} a \right]
\nonumber \\
 &-& \frac{\Delta(t) - \gamma (t)}{2} \left[  a a^{\dag} \rho_S - 2
a^{\dag} \rho_S a + \rho_S a a^{\dag}
 \right], \label{MERWA}
\end{eqnarray}
where we have introduced the bosonic annihilation and creation
operators $a$ and $a^{\dag}$, respectively. The coefficients
$\Delta(t)$ and $\gamma(t)$ are the diffusion and dissipation
coefficients, and they are defined in terms of the noise and
dissipation kernels, respectively \cite{Maniscalco04b}. The form of
Eq. (\ref{MERWA}) is similar to the Lindblad form \cite{Lindblad},
with the only difference that the coefficients appearing in the
master equation are here time dependent. We say that this master
equation is of Lindblad-type when the coefficients $\Delta(t) \pm
\gamma (t)$ are positive at all times \cite{misbelief}. Note,
however, that Lindblad type master equations, contrarily to master
equations of Lindblad form, in general do not satisfy the semigroup
property.

For a high temperature reservoir with Ohmic spectral density the
analytic expression of the diffusion and dissipation coefficients
is the following
\begin{eqnarray}
\Delta(t) &=& 2 g^2 k T \frac{r^2}{1+r^2} \left\{ 1 - e^{-\omega_c
t} \left[ \cos (\omega_0 t) -  (1/r)  \sin (\omega_0 t )\right]
\right\}, \label{deltaHT}
\end{eqnarray}
and
\begin{equation}
\gamma (t)= \frac{g^2 \omega_0 r^2}{1+r^2} \Big[1 \!-\! e^{-
\omega_c t} \cos(\omega_0 t) \! - r e^{- \omega_c t} \sin( \omega_0
t ) \Big], \label{gammasecord}
\end{equation}
with $g$ system-reservoir coupling constant, $r=\omega_c/\omega_0$,
$\omega_c$ cutoff frequency of the spectrum of the reservoir,
$\omega_0$ frequency of the system oscillator, and $k$ Boltzmann
constant.

The solution for the density matrix of the system may be written
in terms of the simmetrically ordered characteristic function
$\chi_t(\xi)$ at time $t$, also known as quantum characteristic
function (QCF), defined through the equation \cite{barnettbook}
\begin{equation}
\label{sdef} \rho_S(t)=\frac{1}{2\pi}\int \chi_t(\xi)\:
e^{\left(\xi a^{\dag}-\xi^* a\right)} d^2\xi.
\end{equation}
Using Eq. (\ref{sdef}) one may solve the master equation given by
Eq. (\ref{MERWA}) by means of a method based on the algebraic
properties of the time evolution superoperator. Indeed we have
proved that the form of the time evolution superoperator may be
factorized into three parts allowing the derivation of an exact
analytic solution \cite{PRAsolanalitica}. The QCF of the system
considered here is given by
\begin{equation}
\chi_t (\xi)=e^{- \Delta_{\Gamma}(t) |\xi|^2} \chi_0 \left[ e^{-
\Gamma (t)/2} e^{-i \omega_0 t} \xi  \right], \label{chit}
\end{equation}
with $\chi_0$ QCF of the initial state of the system. The
quantities $\Delta_{\Gamma}(t)$ and $\Gamma(t)$ appearing in
Eq.~(\ref{chit})
 are defined in terms of the diffusion and dissipation
 coefficients $\Delta(t)$ and $\gamma(t)$ respectively  as follows
\begin{eqnarray}
\Gamma(t)&=& 2\int_0^t \gamma(t_1)\:dt_1, \label{Gamma} \\
\Delta_{\Gamma}(t) &=& e^{-\Gamma(t)}\int_0^t
e^{\Gamma(t_1)}\Delta(t_1)dt_1 \label{DeltaGamma}.
\end{eqnarray}

It is well known that non-Markovian features usually occur in the
dynamics for times $t \ll \tau_R = 2 \pi/ \omega_c$, with $\tau_R$
reservoir correlation time. For quantum optical systems, since
$\omega_c \gg \omega_0$, and $\omega_0$ is an optical frequency,
deviations from the Markovian dynamics appear for very short times,
usually too short to be measured in the experiments. With trapped
ion systems $\omega_0 \simeq 10^{7}$Hz, and therefore non-Markovian
features would appear for times $t \ll 0.1 \mu$s. This is the reason
why typical trapped ion experiments, wherein the time scales go from
$1$ to $100 \mu$s, show a Markovian behaviour. In the trapped ion
context, however, it is possible to engineer an artificial reservoir
wherein parameters as the system-reservoir coupling and the cutoff
frequency can be controlled \cite{engineerNIST}. With an appropriate
engineered reservoir one may force non-Markovian features to appear
by \lq detuning\rq~the trap frequency  from the reservoir spectral
density \cite{Maniscalco04a}. This corresponds, for example, to the
case in which $r=\omega_c / \omega_0 \ll 1$, e.g. $r=0.1$. In this
case the reservoir correlation time is bigger than the period of
oscillation of the ion, $\tau_R = 2 \pi \mu$s, and therefore the
non-Markovian features show up in the time evolution and can be
measured. At the moment, in fact, this would require a slight
modification of the current experimental apparatus used in the group
of David Wineland at NIST since a switch should be positioned
between the filters, which are inside the vacuum chamber, and the
trap electrode \cite{WinelandPV}.

\begin{figure}
\centering
\includegraphics[width=8 cm,height=6 cm]{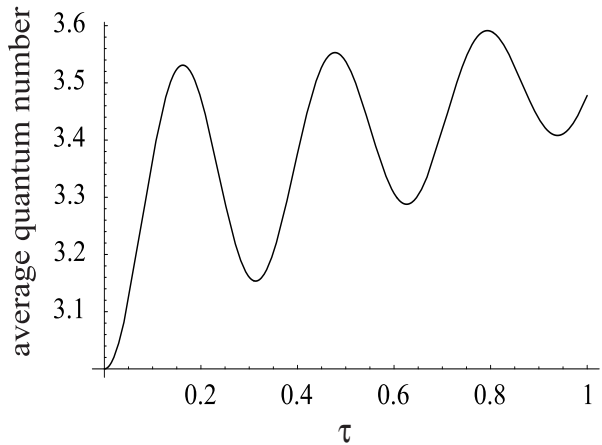}
 \caption{ Time evolution of the average quantum number $\langle n \rangle$ vs the dimensionless time $\tau=\omega_c t$ for
 an initial state having excitation number  $\langle n  \rangle (t=0)= 3$.
 The parameters used are the following $r=0.05$, $g=0.1$, $\omega_c/2 \pi KT =3\cdot10^{-5}$}.\label{fig:0}
\end{figure}

In the following we will pay special attention to the experimentally
realizable regime $r \ll 1$. It is easy to verify that for $r\ll 1$
the coefficients $\Delta(t) \pm \gamma(t)$ appearing in the master
equation, with $\Delta(t)$ and $\gamma (t)$ given by Eqs.
(\ref{deltaHT})-(\ref{gammasecord}), take negative values [see, e.g.
Ref. \cite{misbelief}]. In this case, therefore, the master equation
is of non-Lindblad-type. It is worth underlining that, in this
regime and for the system here considered, the density matrix is
completely positive at all times $t$ while the semigroup property of
the generator of the dynamics is clearly violated
\cite{petruccionebook}. Under these conditions virtual exchanges of
energy between system and reservoir appear  due to the
system-reservoir correlations. These virtual processes show up in
the dynamics of the mean energy of the system in the form of
oscillations (see Fig. \ref{fig:0}), and they are a clear signature
of the break down of the semigroup property. The virtual exchanges
of excitations characterize the regime $r < 1$ since in this regime
the period $T = 2 \pi / \omega_0$ of the system oscillator is
smaller than the reservoir correlation time $\tau_R = 2 \pi /
\omega_c$. Due to the time-energy uncertainty principle, for times
$t<T$, a virtual process consisting in the absorption of a quantum
of energy $\hbar \omega_0$ from a high temperature reservoir and
consequent re-emission of the same quantum of energy may occur. In
Fig. \ref{fig:0} we see the oscillation in the mean number of quanta
$\langle n \rangle$ of the system oscillator for an initial state
having $\langle n \rangle (t=0) = 3$, during the time interval
$0<\tau<1$, with $\tau=\omega_c t$, for $r=0.05$. Note that the
period of the system oscillator, in the same time units, is
$\omega_c T= 2\pi r \simeq 0.3141$, which coincides exactly with the
period of the oscillations displayed in Fig.\ref{fig:0}.

In the following section we will see how virtual processes influence
the dynamics of the Wigner function.

\section{The time evolution of the Wigner function}\label{sec:wigner}
The Wigner function is the Fourier transform of the simmetrically
ordered characteristic function
\begin{equation}
W(\alpha) = \frac{1}{\pi^2} \int_{-\infty}^{\infty} d^2 \xi
\chi(\xi) e^{\alpha \xi^* - \alpha^* \xi}. \label{eq:wfourier}
\end{equation}
Inserting Eq.(\ref{chit}) into the previous equation one obtains
\begin{eqnarray}
W_{t}(\alpha)=\frac{1}{\pi^2}\int_{-\infty}^{\infty}\!\!\!\! d^2 \xi
\, e^{- \Delta_{\Gamma}(t)|\xi|^2} \, e^{\alpha \xi^* - \alpha^*
\xi} \chi_0 ( e^{- \Gamma (t)/2} e^{-i \omega_0 t} \xi). \label{w1}
\end{eqnarray}
Inserting the inverse Fourier transform of Eq. (\ref{eq:wfourier})
into Eq.(\ref{w1}) gives
\begin{eqnarray}
&& W_{t}(\alpha)= \frac{1}{\pi^2} \int_{-\infty}^{\infty} \, d^2
\alpha_0 W_0(\alpha_0) \nonumber \\&&  \int_{-\infty}^{\infty} \,
d^2\xi \, e^{- \Delta_{\Gamma}(t)|\xi|^2} \, e^{b(t, \alpha ,
\alpha_0)
\xi^* - b^*(t, \alpha , \alpha_0) \xi} \nonumber \\
&&= \frac{1}{\pi} \int_{-\infty}^{\infty}  d^2 \alpha_0
W_0(\alpha_0) \frac{ \exp  \left[  - \frac{|b(t, \alpha , \alpha_0
)|^2}{
\Delta_{\Gamma}(t)} \right] }{ \Delta_{\Gamma}(t)} \nonumber \\
&&\equiv \frac{1}{\pi} \int_{-\infty}^{\infty}  d^2 \alpha_0
W_{t}(\alpha|\alpha_0)  W_0(\alpha_0), \label{wfinale}
\end{eqnarray}
with
\begin{equation}
b(t, \alpha , \alpha_0) = \alpha - \alpha_0  e^{- \Gamma (t)/2}
e^{i \omega_0 t}.
\end{equation}
In the derivation of Eq. (\ref{wfinale}) we have used the property
that the Fourier transform of a Gaussian is a Gaussian. The
quantity $W_{t}(\alpha|\alpha_0)$ is the propagator which, for $t
\rightarrow 0$, tends to the delta function $\delta (\alpha -
\alpha_0)$.

If the state of the input field is a coherent state
$|\alpha_0\rangle$, then the integral appearing in Eq.
(\ref{wfinale}) may be easily calculated since the Wigner function
of the initial coherent state is a Gaussian. In this case the Wigner
function at time $t$ reads as follows
\begin{eqnarray}
W_t (\alpha) = \frac{1}{\pi \left[ \Delta_{\Gamma} (t) +
1/2\right]} \exp \left[ \frac{\left| \alpha_0 e^{-\Gamma(t)/2}
e^{-i \omega_0 t} - \alpha \right|^2}{\Delta_{\Gamma} (t) +
1/2}\right].
\end{eqnarray}
Having in mind this equation, and with the help of Eqs.
(\ref{Gamma})-(\ref{DeltaGamma}) and
(\ref{deltaHT})-(\ref{gammasecord}), it is possible to show that the
system-reservoir interaction  spreads the initial Wigner function.
Breathing of the Wigner function, that is the oscillation in its
spread, appears in correspondence of the virtual processes. This is
a dynamical feature which is absent both in the Markovian dynamics
of the damped harmonic oscillator and in the Lindblad-type
non-Markovian dynamics. Indeed, in both the previous regimes, the
spread in the Wigner function simply increases, linearly in time in
the Markovian case, and quadratically in time in the non-Markovian
Lindblad-type case.

We now consider the case of an initially squeezed state. The
initial QCF for squeezed coherent state is
\begin{equation}
\chi_0(\xi)= \exp\left[ -\frac{1}{2} |\xi
C_s-\xi^*e^{-i\phi}S_s|^2 +i(\xi^*\alpha_0^*+\xi\alpha_0)\right].
\end{equation}
Here $C_s=\cosh(s)$ and $S_s=\sinh (s)$, $\alpha_0$ is the
displacement of the initial state of the oscillator and $z=s
e^{-i\phi}$ is the squeezing argument. For this initial condition
the explicit integration appearing in Eq.(\ref{wfinale}) is more
difficult to carry out. In the case of an initial squeezed vacuum
state ($\alpha_0=0$) with squeezing angle $\phi=0$,  the Wigner
function at time $t$ has the form
\begin{eqnarray}
W_t(\alpha)&=& \frac{1}{\pi^2} \int_{-\infty}^{\infty} \, d^2 \xi
\, e^{-\Delta_{\Gamma}(t) |\xi|^2}  e^{\left(
\alpha\xi^*-\alpha^*\xi\right)} \nonumber \\
&& \exp\left[ -\frac{1}{2}e^{-\Gamma(t)} |e^{-i\omega_0 t}\xi C_s
- e^{i\omega_0 t}\xi^*S_s|^2\right].
\end{eqnarray}
Looking at the previous equation one realizes that what we need to
calculate is the Fourier transform of the product of two functions
\begin{equation}
W_t(\alpha)= \frac{1}{\pi^2} \int_{-\infty}^{\infty} \, d^2 \xi \,
f_1(\xi)\, f_2(\xi) e^{\left( \alpha\xi^*-\alpha^*\xi\right)},
\end{equation}
where
\begin{eqnarray}
f_1(\xi) &=& e^{-\Delta_{\Gamma}(t) |\xi|^2}, \\
f_2(\xi) &=& \exp\left[ -\frac{1}{2} e^{-\Gamma(t)} |e^{-i\omega_0
t}\xi C_s - e^{i\omega_0 t}\xi^*S_s|^2\right].
\end{eqnarray}

\begin{figure}
\centering
\includegraphics[width=8 cm,height=6 cm]{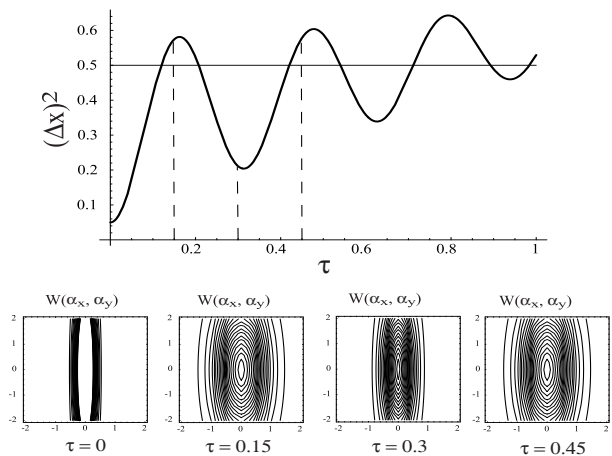}
 \caption{ Time evolution of  $(\Delta x)^2$ vs the dimensionless time $\tau=\omega_c t$. The parameters used are the following
 $\sigma^2=0.1$, $r=0.05$, $g=0.1$, $\omega_c/2 \pi KT = 3\cdot10^{-5}$. The insets at the bottom of the figure show the
 contour plot of the Wigner function at times $\tau=0$, $\tau=0.15$, $\tau=0.3$, $\tau=0.45$.}\label{fig:1}
%\end{centering}
\end{figure}

If we denote the Fourier transform $\mathcal{F}$ of the functions
$f_1(\xi)$ and $f_2(\xi)$ as follows, $F_1=\mathcal{F}(f_1)$ and
$F_2=\mathcal{F}(f_2)$, we can recast $W_t$ in the form
\begin{equation}
W_t(\alpha) = F_1 \ast F_2
\end{equation}
where $\ast$ denotes the convolution. We now extend the method
developed in Ref.~\cite{Matsuo93} for the calculation of the
Wigner function of an initial squeezed state to the non-Markovian
case we are dealing with in this paper. Carrying out the
calculations, in a frame rotating with $\omega_0$, one obtains
\begin{eqnarray}
W_{t}(\alpha) &=& M \exp\left[ \frac{-2
\alpha_x^2}{2\Delta_{\Gamma}(t)+
e^{-\Gamma(t)}(C_{2s}+S_{2s})^{-1}}\right]\nonumber
\\ &\times& \exp\left[ \frac{-2 \alpha_y^2}{2\Delta_{\Gamma}(t)+
e^{-\Gamma(t)}(C_{2s}-S_{2s})^{-1}}\right] \nonumber \\
&=& M \exp \left[\frac{- \alpha_x^2}{(\Delta x)^2(t)} + \frac{-
\alpha_y^2}{(\Delta y)^2(t)} \right] \label{eq:wignersqueez}
\end{eqnarray}
Here, $\alpha_x$ and $\alpha_y$ are the real and imaginary parts
of $\alpha$, $M$ is a time dependent normalization constant, and
\begin{eqnarray}
(\Delta x)^2 (t)=\Delta_{\Gamma}(t)+
\frac{e^{-\Gamma(t)} e^{-2s}}{2}, \\
(\Delta y)^2 (t)=\Delta_{\Gamma}(t)+ \frac{e^{-\Gamma(t)}
e^{2s}}{2},
\end{eqnarray}
are the variances of the dimensionless quadratures
\begin{eqnarray}
x &=& \frac{1}{\sqrt{2}} \left( a + a^{\dag} \right), \label{eq:x}\\
y &=& \frac{-i}{\sqrt{2}} \left( a - a^{\dag} \right).
\label{eq:y}
\end{eqnarray}
At $t=0$, $(\Delta x)^2 (0)= e^{-2s}/2 \equiv \sigma^2/2$ and
$(\Delta y)^2 (0)= e^{2s}/2 \equiv 1/2 \sigma^2$. We consider an
initial state displaying squeezing in the $x$ quadrature, i.e., in
the trapped ion case, a state squeezed along the $x$ direction of
the trap. This corresponds to $\sigma < 1$ since the $x$-squeezing
is characterized by $(\Delta x)^2<0.5$. In Fig. \ref{fig:1} we plot
the time evolution of $(\Delta x)^2$ as a function of the
dimensionless time $\tau=\omega_c t$, for an initial squeezed state
with $\sigma^2=0.1$, together with the contour plot of the Wigner
function at four different time instants, $\tau=0$, $\tau=0.15$,
$\tau=0.3$ and $\tau=0.45$.
The graphic shows that, due to the interaction with a high
temperature reservoir, the variance of the $x$ dimensionless
operator, defined by Eq.(\ref{eq:x}), increases from its initial
value $(\Delta x)^2 (0)= 0.05$. For short times, however, the
dynamics is governed by oscillations between a squeezed state and a
non-squeezed state due to the occurrence of virtual exchanges of
excitations between the system oscillator and the reservoir. As
clearly shown in the figure, at $\tau \simeq 0$ and $\tau \simeq
0.15$ the system is in a squeezed state [$(\Delta x)^2 (0)< 0.5$]
while at $\tau \simeq 0.3$ and $\tau \simeq 0.45$ the system is in a
non-squeezed state. By using the calculation we have carried out in
this paper for the time evolution of an initially squeezed state,
see Eq. (\ref{eq:wignersqueez}), we can also plot the contours of
the Wigner function at these time instants [see insets at the bottom
of of Fig. (\ref{fig:1})]. More in  general Eq.
(\ref{eq:wignersqueez}), together with Eqs.
(\ref{Gamma})-(\ref{DeltaGamma}) and
(\ref{deltaHT})-(\ref{gammasecord}), tell us that the Wigner
function of an initially squeezed state, in the phase space, spread
along both the $\alpha_x$ and the $\alpha_y$ axes and eventually
approaches the Wigner function of a thermal state. For times shorter
than the reservoir correlation time, however, in the non-Lindblad
regime both $(\Delta x)^2$ and $(\Delta y)^2$ oscillate, so that if
the system is initially in a squeezed state along the $x$ ($y$)
direction then this may result in $x$ ($y$) squeezing oscillations.

\section{Conclusions}\label{sec:conclusions}
In this paper I have studied the non-Markovian dynamics of a quantum
harmonic oscillator in phase space. I have presented a method to
calculate the Wigner function of an initially squeezed state of
motion for short times, when the system-reservoir correlations
dominate the dynamics. The solution obtained is used to investigate
the non-Lindblad regime, characterized by virtual exchanges of
energy between the system and the reservoir. It is shown that the
virtual processes are responsible for the oscillations between a
squeezed and a non-squeezed state of the system. It is worth
recalling that the non-Lindblad regime is experimentally realizable
with artificial reservoirs in the trapped ion context. In this
context, moreover, the Wigner function is experimentally measurable.
Therefore, by measuring the Wigner function, it is possible to
reveal the virtual processes characterizing the non-Markovian
non-Lindblad dynamics, and in particular the squeezing-nonsqueezing
oscillations. The result here presented contributes to the
understanding of the short time dynamics of an ubiquitous model of
an open quantum system, namely the damped harmonic oscillator, and
hence it is also of potential interest for quantum technologies. In
fact in many solid-state physical systems promising for building
prototypes of quantum devises the Markovian approximation does not
hold. Moreover, if one is interested to very short system dynamics
one needs to abandon the Markovian assumption. The result has also
implications in the study of fundamental issues of quantum
measurement theory since there are indications that the non-Lindblad
regime may correspond to a situation in which no measurement scheme
interpretation of the non-Markovian stochastic process exists.

\section*{Acknowledgements}
The author thanks Jyrki Piilo for stimulating discussions on the
subject of the paper. Financial support from the EU network CAMEL
(contract MTKD-CT-2004-014427) and from the Italian National
Foundation Angelo Della Riccia is gratefully acknowledged.


\section*{References}

\begin{thebibliography}{99}

\bibitem{nielsen}  Nielsen M A and  Chuang I L 2000 {\it Quantum Computation and
Quantum Information} (Cambridge: Cambridge University Press).

\bibitem{petruccionebook}  Breuer H-P and  Petruccione F 2002 {\it The Theory of Open
Quantum systems}  (Oxford: Oxford University Press) .


\bibitem{Alicki02}
 Alicki R,  Horodecki M, Horodecki P and
Horodecki R 2002  {\it Phys. Rev. A} \textbf{65} 062101; Alicki R
 {\it et al} 2004  {\it Phys. Rev. A}
\textbf{70} 010501.

\bibitem{fotolaser}   Hope J J 1997 {\it Phys. Rev.
A} \textbf{55} R2531;  Moy G M,  Hope J J, and  Savage C M 1999 {\it
Phys. Rev. A} \textbf{59} 66;  John S and  Quang T 1994 {\it Phys.
Rev. Lett.} \textbf{74} 3419.

\bibitem{QBMpapers}
 Haake F and  Reibold R 1985 {\it Phys. Rev. A} {\bf 32} 2462;
 Feynman R P and  Vernon F L 1963 {\it Ann. Phys.} {\bf 24} 118;
 Hu B L, Paz P, and  Zhang Y 1992 {\it Phys. Rev. D} {\bf 45} 2843;
Ford G W and  O'Connell R F 2001 {\it Phys. Rev. D} {\bf 64} 105020;
 Grabert H,  Schramm P, and  Ingold G-L 1988 {\it Phys. Rep.} {\bf 168}
 115.


\bibitem{PRAsolanalitica} Intravaia F,  Maniscalco S, and
Messina A 2003 {\it Phys. Rev. A} {\bf 67} 042108.

\bibitem{Maniscalco04b}
Maniscalco S,  Piilo J, Intravaia F, Petruccione F, and  Messina A
2004 {\it Phys. Rev. A} {\bf 70} 032113.

\bibitem{Maniscalco04a}
Maniscalco S,  Piilo J, Intravaia F, Petruccione F, and  Messina A
2004 {\it Phys. Rev. A} {\bf 69} 052101.

\bibitem{Monroe96} Leibfried D. {\it et al.} 1996
{\it Phys. Rev. Lett. 77} 4281.



\bibitem{EPJRWA} For a detailed discussion  on the approximations leading to this master equation see Intravaia F,  Maniscalco S, and
Messina A 2003 {\it Eur. Phys. J. B} {\bf 32} 109.

\bibitem{Lindblad}
Lindblad G 1976 {\it Comm. Math. Phys.} {\bf 48} 119; Gorini V,
Kossakowski A, and  Sudarshan E C G 1976 {\it J. Math. Phys.} {\bf
17} 821.

\bibitem{misbelief}  Maniscalco S,  Intravaia F,  Piilo J, and  Messina A 2004  {\it J. Opt. B:
Quantum and Semiclass. Opt.} {\bf 6} S98.

\bibitem{barnettbook}
Barnett S M and Radmore P M 1997 {\it Methods in Theoretical Quantum
Optics} (Oxford: Oxford University Press).





\bibitem{engineerNIST}
 Myatt C J {\it et al.} 2000 {\it Nature} {\bf 403} 269;
Turchette  Q A {\it et al.} 2000 {\it Phys. Rev. A} {\bf 62} 053807.

\bibitem{WinelandPV} David Wineland, private communication.

\bibitem{Matsuo93}
Matsuo K 1993 {\it Phys. Rev. A} {\bf 47} 3337.

\end{thebibliography}
\end{document}